\documentclass[journal]{IEEEtran}

\IEEEoverridecommandlockouts

\usepackage{tikz}
\usepackage{textcomp}
\usepackage{lipsum}

\newcommand\copyrighttext{%
  \footnotesize
  \textit{N. Pompeo, H. Schneidewind, E. Silva, IEEE Trans. Appl. Supercond., accepted for publication (2019)}  \vspace{.2cm}\\
  \textcopyright 2019 IEEE. Personal use of this material is permitted.
  Permission from IEEE must be obtained for all other uses, in any current or future
  media, including reprinting/republishing this material for advertising or promotional
  purposes, creating new collective works, for resale or redistribution to servers or
  lists, or reuse of any copyrighted component of this work in other works.
  }
\newcommand\copyrightnotice{%
\begin{tikzpicture}[remember picture,overlay]
\node[anchor=north,yshift=-10pt] at (current page.north) {\fbox{\parbox{\dimexpr\textwidth-\fboxsep-\fboxrule\relax}{\copyrighttext}}};
\end{tikzpicture}%
}

\ifCLASSINFOpdf

\else
  \fi

\usepackage{graphicx}
\usepackage{amsmath}
\usepackage{amssymb}
\usepackage{mathrsfs}

\newcommand{\rmi}{ \mathrm{i} }

\usepackage{color}
\usepackage{xspace}
\usepackage{color}
\usepackage[normalem]{ulem}

\newcommand{\textoutr}[1]{}

\begin{document}

\newcommand{\IJMPB}{Int. J. Mod. Phys. B }
\newcommand{\PC}{Physica C }
\newcommand{\PB}{Physica B }
\newcommand{\JS}{J. Supercond. }
\newcommand{\IEEEmw}{IEEE Trans. Microwave Theory Tech. }
\newcommand{\IEEEas}{IEEE Trans. Appl. Supercond.}
\newcommand{\IEEEim}{IEEE Trans. Instr. Meas. }
\newcommand{\PRB}{{Phys. Rev. B}}
\newcommand{\PRL}{{Phys. Rev. Lett.}}
\newcommand{\PR}{Phys. Rev. }
\newcommand{\PL}{Phys. Lett. }
\newcommand{\IJIMW}{Int. J. Infrared Millim. Waves }
\newcommand{\APL}{{Appl. Phys. Lett. }}
\newcommand{\JAP}{{J. Appl. Phys. }}
\newcommand{\JPCM}{J. Phys.: Condens. Matter }
\newcommand{\JPCS}{J. Phys. Chem. Solids }
\newcommand{\AdP}{Adv. Phys. }
\newcommand{\Nat}{Nature }
\newcommand{\CM}{cond-mat/ }
\newcommand{\JpnJAP}{Jpn. J. Appl. Phys. }
\newcommand{\PhT}{Phys. Today }
\newcommand{\ZETF}{Zh. Eksperim. i. Teor. Fiz. }
\newcommand{\JETP}{Soviet Phys.--JETP }
\newcommand{\EL}{Europhys. Lett. }
\newcommand{\Sci}{Science }
\newcommand{\EJPB}{Eur. J. Phys. B }
\newcommand{\IJMB}{Int. J. of Mod. Phys. B }
\newcommand{\RPP}{Rep. Prog. Phys. }
\newcommand{\SUST}{Supercond. Sci. Technol. }
\newcommand{\JLTP}{J. Low Temp. Phys. }
\newcommand{\RSI}{Rev. Sci. Instrum. }
\newcommand{\RMP}{Rev. Mod. Phys. }
\newcommand{\LTP}{Low Temp. Phys. }
\newcommand{\etal}{{\it et al}}

\title{Measurements of microwave vortex response in dc magnetic fields in Tl$_2$Ba$_2$CaCu$_2$O$_{8+x}$ films}

\author{N.~Pompeo,~\IEEEmembership{Senior Member,~IEEE,}
        H. ~Schneidewind,
        E.~Silva,~\IEEEmembership{Senior Member,~IEEE}%
\thanks{E. Silva and N. Pompeo are with the Department
of Engineering, Universit\`{a} Roma Tre, 00146 Roma,
Italy. Corresponding author: E. Silva; e-mail: enrico.silva@uniroma3.it.}%
\thanks{H. Schniedewind is with the Leibniz Institute of Photonic Technology, Albert-Einstein-Stra\ss e 9, D-07745 Jena, Germany.}%
\thanks{Manuscript received October 30, 2018.}}

{}

\maketitle
\copyrightnotice

\begin{abstract}
There is a renewed interest in superconductors for high-frequency applications, leading to a reconsideration of already known low-$T_c$ and high-$T_c$ materials. In this view, we present an experimental investigation of the millimeter-wave response in moderate magnetic fields of Tl$_2$Ba$_2$CaCu$_2$O$_{8+x}$ superconducting films with the aim of identifying the mechanisms of the vortex-motion-induced response. We measure the dc magnetic-field-dependent change of the surface impedance, $\Delta Z_s(H) = \Delta R_s(H) + i\Delta X_s(H)$ at 48 GHz by means of the dielectric resonator method. We find that the overall response is made up of several contributions, with different weights depending on the temperature and field: a possible contribution from Josephson or Abrikosov-Josephson fluxons at low fields; a seemingly conventional vortex dynamics at higher fields; a significant pair breaking in the temperature region close to $T_c$. We extract the vortex motion depinning frequency $f_p$, which attains surprisingly high values. However, by exploiting the generalized model for relaxational dynamics we show that this result come from a combination of a pinning constant $k_p$ arising from moderate pinning, and  a vortex viscosity $\eta$ with anomalously small values. This latter fact, implying large dissipation, is likely a result from a peculiar microscopic structure and thus poses severe limits to the application of Tl$_2$Ba$_2$CaCu$_2$O$_{8+x}$  in a magnetic field. 
\end{abstract}


\IEEEpeerreviewmaketitle

\section{Introduction}
\label{intro}
\IEEEPARstart{T}{he} need for pushing further the performances of superconductor-based high--frequency devices, such as accelerating cavities \cite{cavities} or electromagnetic screenings \cite{FCC} has given rise to a reexamination of the temperature and magnetic field dependences of the surface impedance in both conventional and high--$T_c$ superconductors (HTS). Well--known superconductors, such as Nb$_3$Sn, are being investigated with respect to the field--dependent surface impedance \cite{andrea}, and NbTi cavities for dark matter search are being tested in mid-to-high magnetic fields \cite{digioacchino}. HTS have been studied for longtime, but Tl-based HTS have been dismissed in early times due to their inferior performances with respect to YBa$_2$Cu$_3$O$_{7-\delta}$ (YBCO). However, some Tl-based compounds are now under scrutiny for selected large--scale applications \cite{vaglio}. It is then a useful contribution to investigate the microwave properties in magnetic field of other superconductors. Clarifying the main mechanisms of the dissipation is the main step in order to ascertain the usefulness of a material for selected applications.

Since the microwave response in a dc magnetic field is of interest here, one has to deal with losses mainly due to fluxon motion \cite{Marcon1991}. However, other sources of dissipation cannot be ruled out: weak--links, and related ``strange fluxons'' such as Josephson or Abrikosov-Josephson fluxons \cite{gureAJ}, and quasiparticle excitations. It should be noted, however, that weak--links phenomena can be reduced or avoided with proper material engineering, and quasiparticle excitation is an intrinsic detrimental effect that arises -for what concerns applications- only close to $T_c$ or $H_{c2}$. It is then useful to give more detail on the main source of dissipation, that is the fluxon motion. The response of the system of flux lines to an ac field is given by the vortex complex resistivity $\rho_{vm}$ \cite{CC,brandt,PompeoPRB08}:
\begin{equation}
\label{eq:rhovm}
    \rho_{vm}=\rho_{vm1}+\rmi\rho_{vm2}=\frac{\Phi_0B}{\eta}\frac{\chi+\rmi\frac{f}{f_{0}}}{1+\rmi\frac{f}{f_{0}}}
\end{equation}
\noindent where $\Phi_{0}\simeq$ 2.07$\cdot$10$^{-15}$ Tm$^2$ is the flux quantum, $\eta$ is the so--called vortex viscosity, $B\simeq\mu_0H$ (in the London limit) is the magnetic induction field, and $\chi$ is a dimensionless creep factor that takes into account the thermal activation phenomena, with $\chi \in [0,1]$. This expression simplifies when no thermal effects are relevant, in which case $\chi =$ 0 and $f_0 \to f_p$, where $f_p=k_p/(2\pi\eta)$ is the so--called depinnning frequency, and one writes down the well--known result by Gittleman and Rosenblum \cite{GR}:
\begin{equation}
\label{eq:rhoGR}
    \rho_{vm}=\frac{\Phi_0B}{\eta}\frac{1}{1-\rmi\frac{f_p}{f}}
\end{equation}
where it is clear that $f_p$ is the crossover frequency that separates the low frequency, vanishing dissipation (Campbell) regime \cite{Campbell}) from the high-frequency dissipative (flux--flow) regime. The depinning frequency has been recognized as a fundamental parameter to assess the microwave properties of superconductors \cite{GR, vaglionup}, since it gives a rough measure of the weight of the losses at a certain frequency: the dimensionless parameter $r=\rho_{vm2}/\rho_{vm1}\simeq f_p/f \lessgtr 1$ is the crossover mark between Campbell and flux--flow regime (one has however to bear in mind that the crossover region extends for a decade in frequency \cite{GR}).

The aim of this paper is to report on the microwave surface impedance of thin Tl$_2$Ba$_2$CaCu$_2$O$_{8+x}$ (TBCCO) films in a moderate magnetic field $\mu_0H \leq 0.8$ T, in order to identify the main dissipative mechanisms and to give information on some of the relevant parameters to assess the high--frequency performances, with particular attention to the depinning frequency and to its dependence on the temperature and magnetic field. We will compare the results to typical results obtained in YBCO and in conventional superconductors.

\begin{figure}[hbt!]
\centerline{\includegraphics[width=0.95\columnwidth]{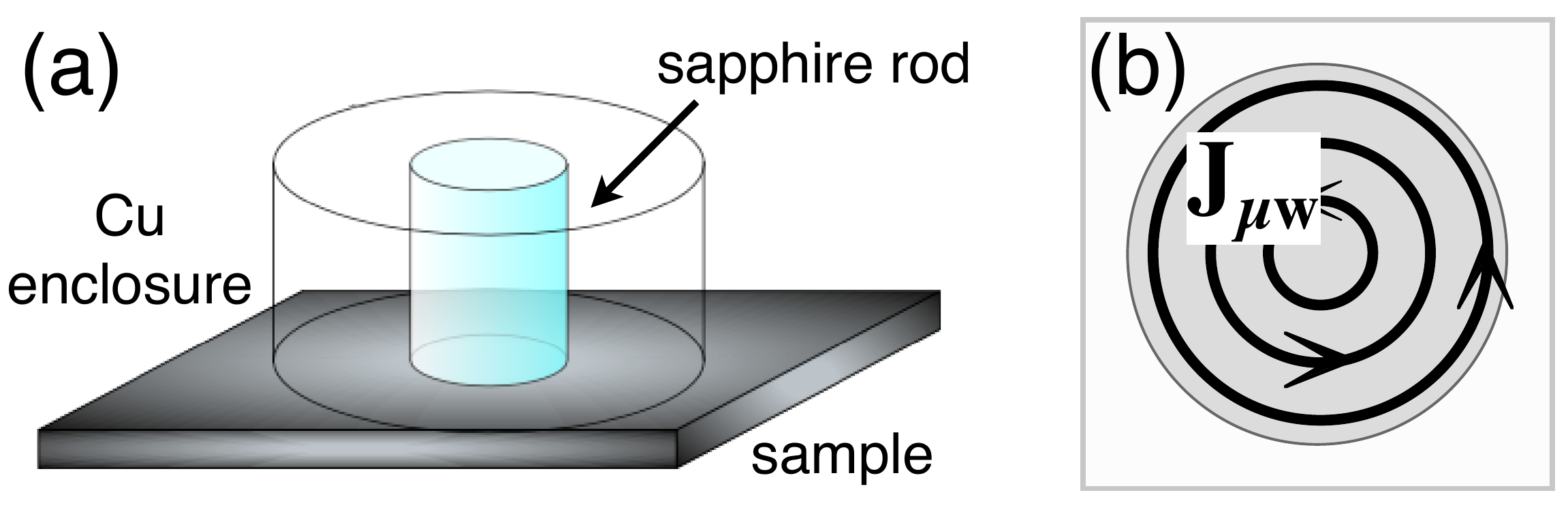}}
  \caption{(a) sketch of the resonator and of the sample placement. (b) sketch of the microwave patterns on the sample surface.}
\label{fig:exp}
\end{figure}

\section{Samples and method}
\label{sec:exp}
Measurements are performed on two TBCCO thin films (thickness $t_s$=240 nm), labelled as TS2 and TS5,  grown by conventional two-step method in the c-axis direction over CeO$_2$ buffered R-plane sapphire substrates, 2 inches in diameter and 0.5~mm thick. Preparation details and additional characterization are available elsewhere \cite{schneidewindIPCF00,schneidewindSUST01}. The samples were cut from a wafer in the shape of $10\times10$~mm$^2$ squares. The film thickness is $t_s\simeq$ 240 nm. Since both samples give analogous results, we report here data obtained on sample TS2, where inductive measurements yielded $T_{c}\simeq$104 K, which compares well to the maximum $T_c\approx$110 K in very clean samples \cite{chuTl}, and $J_{c}$(77~K, $H$=0) = 1.0~MAcm$^{-2}$.

The microwave surface impedance is measured by means of a cylindrical sapphire resonator \cite{PompeoMSR14} used within the surface perturbation method. The resonator is operated in reflection in the TE$_{011}$ mode, with the resonance frequency $f_{res}\approx$47.7~GHz. The measuring frequency is much smaller than the gap, estimated in the tens of meV (40 meV, equivalent to approximately 10 THz, as measured in the infrared range \cite{wangGap}. Microwave patterns induced on the sample are planar and circular, so that only the $ab$ plane response is measured. The unloaded $Q$ factor and the resonance frequency $f_{res}$ yield the surface impedance \cite{PompeoMSR14}. Taking into account the thickness of the film \cite{Pompeo2017b}, one can write down the following relation between the measured $Q,\;f_{res}$, the effective surface impedance shift $\Delta Z_{s}(H)$ and the microwave complex \textit{resistivity} $\tilde\rho$:
\begin{eqnarray}
\label{eq:measrho}
    \Delta Z_{s}(H)\simeq\frac{\Delta\tilde{\rho}(H)}{t_s}=\frac{\Delta{\rho_1}(H)+\rmi\Delta{\rho_2}(H)}{t_s}=\\
    \nonumber
    =G_s\left(\frac{1}{Q(H)}-\frac{1}{Q(0)}-2\rmi\frac{f_{res}(H)-f_{res}(0)}{f_{res}(0)}\right)
\end{eqnarray}
\noindent where $G_s$ is a calculated geometrical factor.

Using a solid/liquid nitrogen cryostat we vary the temperature between $T\approx$ 60 K and $T_c$. A moderate magnetic field $\mu_0H\leq0.8$ T is applied perpendicularly to the sample surface (i.e. parallel to the superconductor $c$-axis).

\begin{figure}[hbt!]
\centerline{\includegraphics[width=0.9\columnwidth]{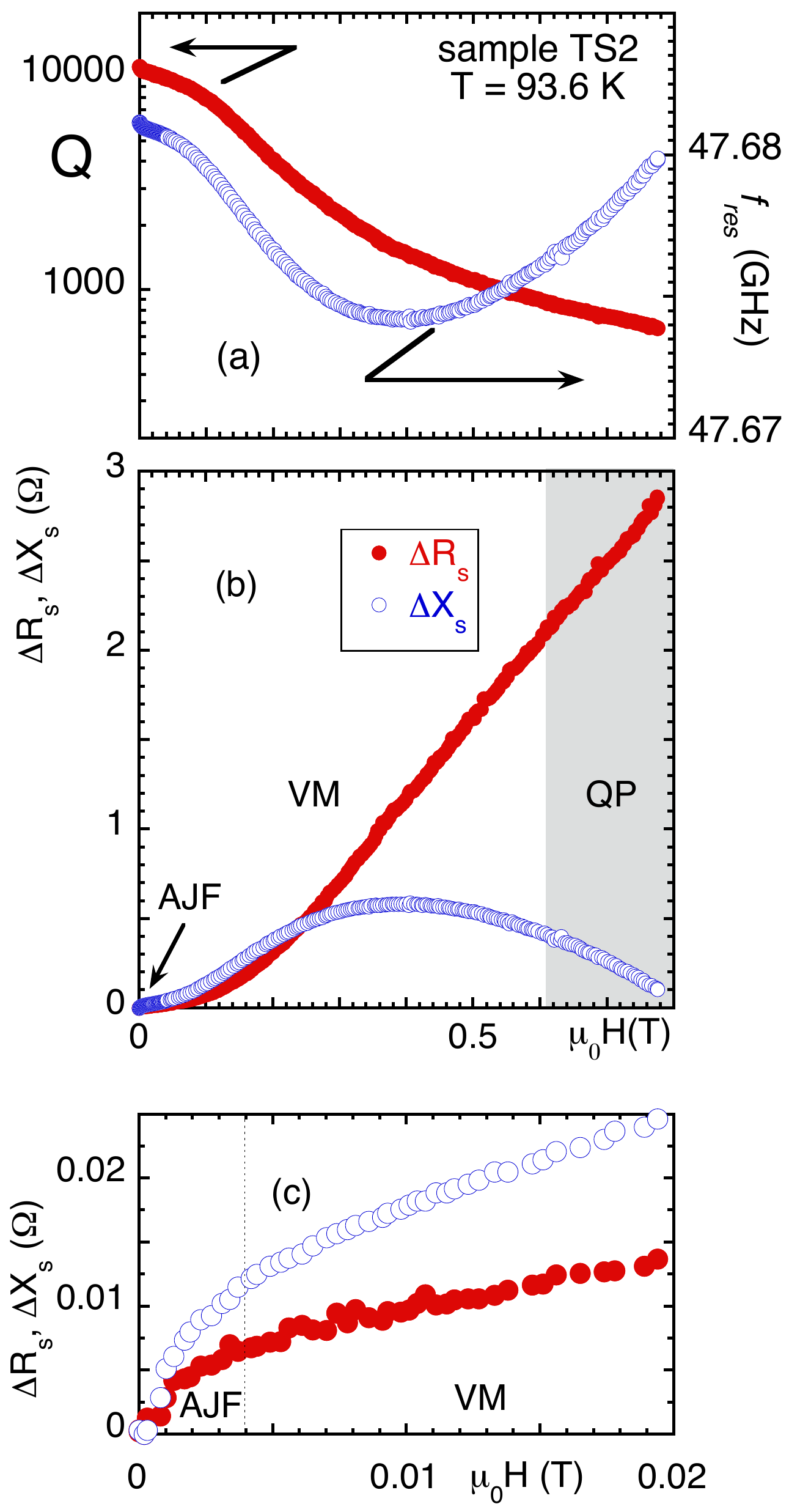}}
  \caption{(a) Field dependence of the quality factor $Q$ (full red symbols) and of the resonant frequency $f_{res}$ (open blue circles). (b) Field induced variations $\Delta R_s =\Delta\rho_1/t_s$ (full red symbols) and $\Delta X_s =\Delta\rho_2/t_s$ (open blue circles), derived from the data in (a) according to Eq.\eqref{eq:measrho}; the grey region indicates  the crossover region towards the QP regime (see text), the arrow points to the AJF field region, invisible on this scale. (c) Enlargement of the data in (b) for the low field region; we indicate the field regions where AJF or VM regimes are dominant.}
\label{fig:data}
\end{figure}

A typical measurement is shown in Figure \ref{fig:data}, where we show the raw data for $Q$ and $f_{res}$ vs. $H$ (Fig.\ref{fig:data}a) and the resulting surface impedance shift  $\Delta\tilde\rho(H)$  (Fig.\ref{fig:data}b). Figure \ref{fig:data}c shows an enlargement of the low--field region, illustrating the wide dynamic range of the measurements.

We briefly comment on Fig.s \ref{fig:data}b and \ref{fig:data}c. The sample data have been selected at a rather high temperature in order to highlight all three regimes that showed up in the measurements. First, at very low fields (Fig.\ref{fig:data}c) a small ``step'' in the surface impedance appears, with a trend to saturation at a few mT. Second (Fig.\ref{fig:data}b), a rapid increase (first linear, then superlinear) of both $\Delta R_{s}$ and $\Delta X_{s}$ is observed, followed by a field region where $\Delta X_{s}$ only drops quickly. For reasons that will be clarified later, we label these regimes as AJF, VM, QP, respectively. We anticipate that the discussion will be focused on the intermediate, VM region. All the regions here depicted have different extension (i.e., field range) depending on the temperature. In particular, at low $T$, below $\sim$85 K, the QP region cannot be observed.

\section{Experimental results and discussion}
\label{sec:results}
\begin{figure}[hbt!]
\centerline{\includegraphics[width=0.7\columnwidth]{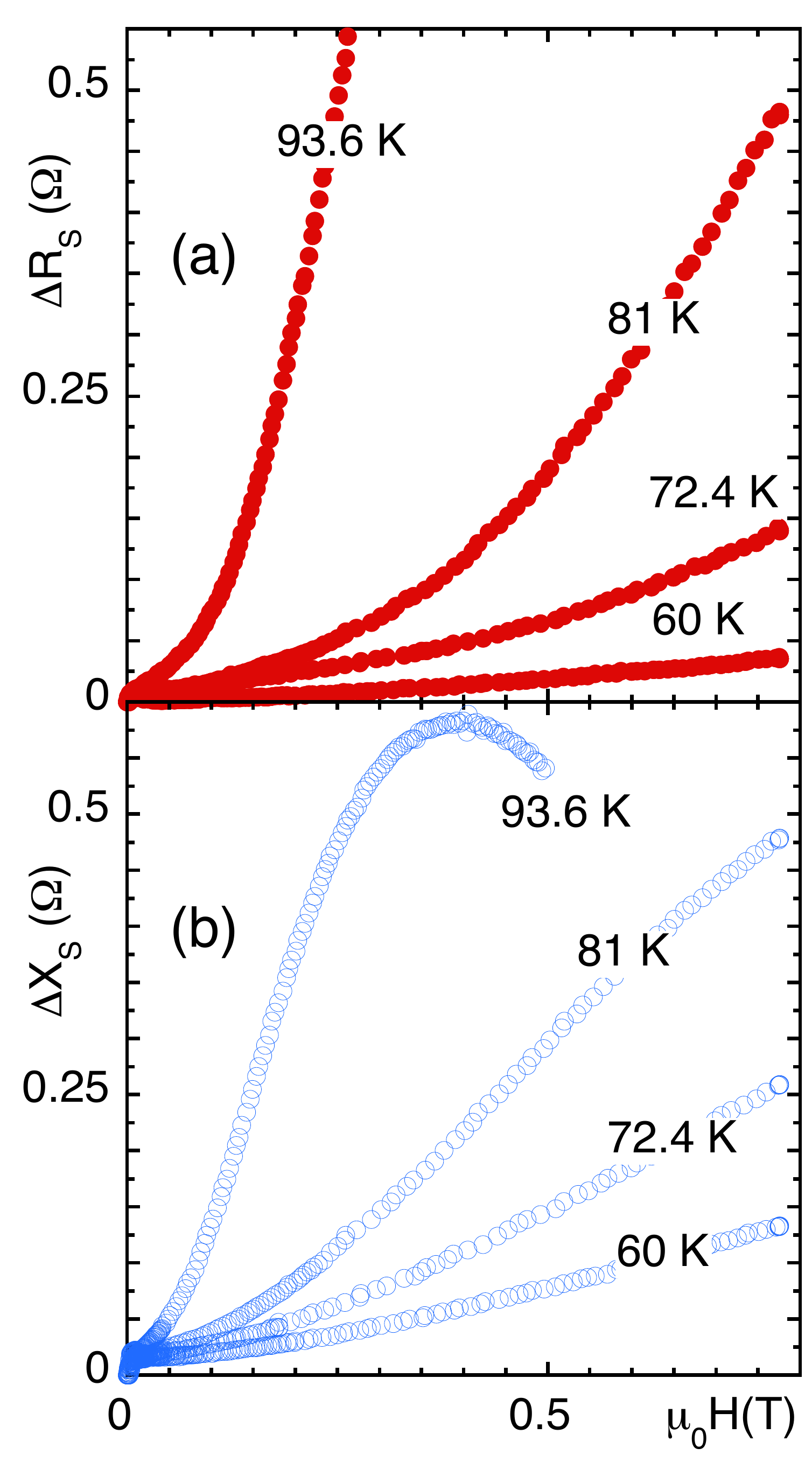}}
  \caption{(a) Field dependence of the surface resistance $\Delta R_S =\Delta\rho_1/t_s$ (full red symbols) in sample TS2 at selected temperatures. (b) Field dependence of the surface resistance $\Delta X_s =\Delta\rho_2/t_s$ (open blue circles) in sample TS2 at selected temperatures. At all temperatures one observes $\Delta X_s > \Delta R_s$.}
\label{fig:DeltaRX}
\end{figure}

We report a series of measurements at selected temperatures in Fig.\ref{fig:DeltaRX}. It can be seen that, at all temperatures, $\Delta X_s >\Delta R_s$. This is usually a manifestation of very strong pinning: the elastic energy stored in the fluxon system is larger than the dissipated energy per oscillation due to the flux--flow mechanism. The customary parameter to quantify the strength of the pinning is, at microwaves, the parameter $r=\Delta X_s /\Delta R_s$. Making reference to Eq.\eqref{eq:rhoGR} and to Eq.\eqref{eq:measrho}, it is seen that, should the field--increase of the surface impedance be due to vortex motion only, one would have $r=\Delta\rho_{vm2}/\Delta\rho_{vm1}$. In the present case, as anticipated in Sec.\ref{sec:exp}, several mechanisms are present so that a preliminary data elaboration is needed.

First, we note that the field region where $\Delta X_s$ decreases with the field is a clear manifestation of quasiparticle (QP) increase (or pair--breaking): remember that, since we are dealing with thin films, the effective measured surface impedance is given by the thin film approximation as per Eq.\eqref{eq:measrho}. In the normal state the imaginary resistivity is clearly zero (at our frequencies), so that the drop to zero in the effective $\Delta X_s$ is a straightforward manifestation of pair--breaking.

Second, the low field region, such as the one reported in Fig.\ref{fig:data}c, shows a behavior reminiscent of the dynamics of Abrikosov-Josephson fluxons (AJF) threading the grain boundaries of the sample \cite{gureAJ,gurePin,gureDyn}. Even a preliminary treatment of this phenomenon is out of the scope of this paper. To our purposes, it is sufficient to observe that another process, most likely due to grain boundaries, adds to the conventional vortex motion and to the pair--breaking. A detailed analysis of the AJF regime is deferred to a future work.

To take into account simultaneously all three processes is rather cumbersome and prone to large uncertainties in the fitting parameters. We adopt a different strategy. We can consider the measured $\Delta Z_s$ as the sum of three different contributions, as $\Delta Z_s = \Delta Z_{vm}+\Delta Z_{AJF}+\Delta Z_{QP}$. We are interested in the derivation of the vortex--motion parameters present in $\Delta Z_{vm}$  through Eq. \eqref{eq:rhoGR}. We then remove the contribution of $\Delta Z_{QP}$ by simply avoiding to examine the data where $\Delta X_s$ decreases with the field (this is why some of the following data will be cut above a certain field). We then remove the $\Delta Z_{AJF}$ contribution by estimating the height of the saturation value of the first, low field increase of $\Delta Z_s$ (as an example, in Fig.\ref{fig:data}c we estimate $\Delta Z_{AJF}\simeq 8\;\mathrm{m\Omega} +\rmi 12\;\mathrm{m\Omega}$), and we subtract it from the data. We note however that $\Delta Z_{AJF}$ is negligible with respect to the overall $\Delta Z_s$, so that we expect small additional errors in the derivation of the vortex motion (VM) parameters. After the subtraction, and the field limitation if required, we analyse the data with Eq. \eqref{eq:rhoGR}.  A consequence of the need to remove the other contributions is that the parameters obtained from the analysis are not reliable below $\sim$0.2 T, so that the following plots will report the data in the range 0.2 T -- 0.8 T only.

We use the procedure described with great detail in \cite{PompeoPRB08} to derive the vortex parameters. Here, the parameters are derived from the data in Fig.\ref{fig:DeltaRX} by using Eq.\eqref{eq:rhoGR}. We refer the reader to \cite{PompeoPRB08} for the estimates of the uncertainties and of the statistical confidence intervals.

\begin{figure}[hbt!]
\centerline{\includegraphics[width=0.85\columnwidth]{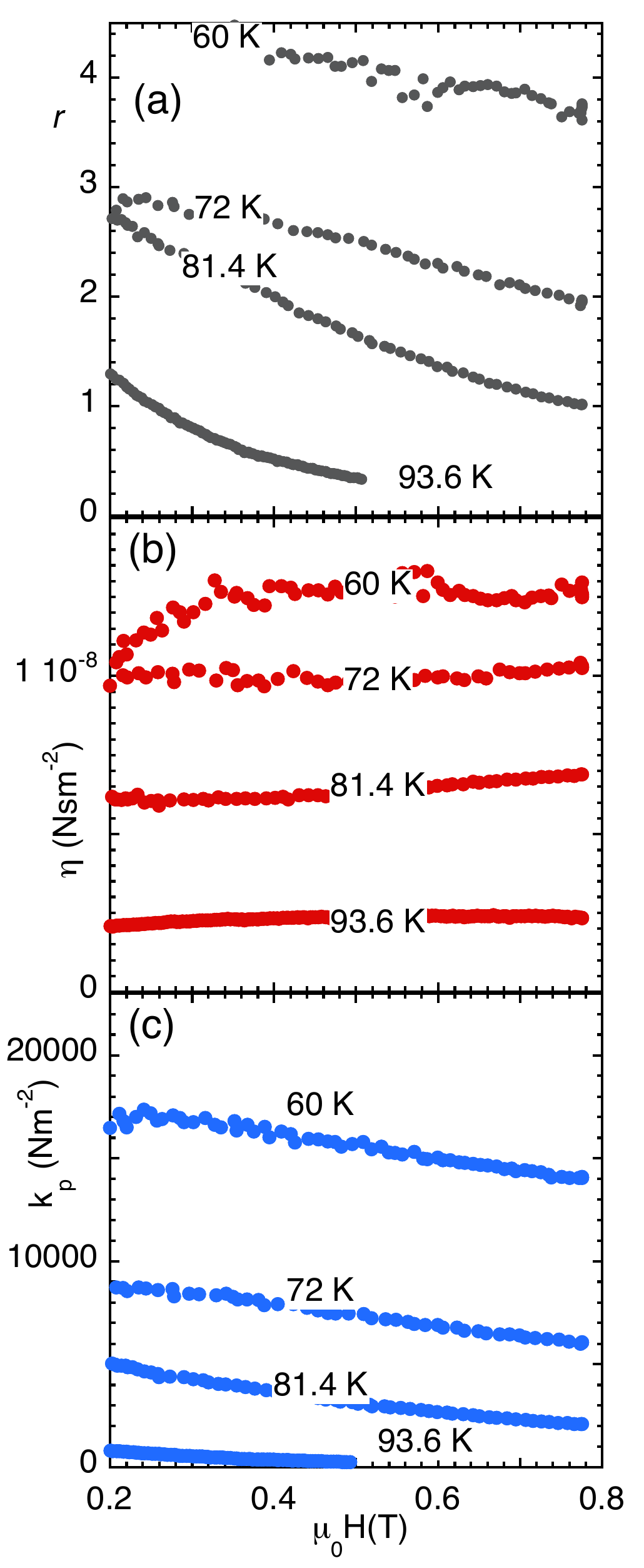}}
  \caption{(a) Pinning parameter $r=\Delta\rho_{vm2}/\Delta\rho_{vm1}$. (b) Vortex viscosity $\eta$. (c) Pinning constant $k_p$.}
\label{fig:param}
\end{figure}

The first important parameter is the depinning frequency $f_p$. We report such data in the form $r=\Delta\rho_{vm2}/\Delta \rho_{vm1}$ in Figure \ref{fig:param}a. As it can be seen, the data show very high values $r > 1$. Bearing in mind our operating frequency, $f\simeq$ 48~GHz, our values for $r$ would give a huge $f_p\approx$ 200~GHz at low temperature. This is a surprisingly large value:  the depinning frequency attains values of $\sim$10~GHz (at $\sim T_c/2$) in YBa$_2$Cu$_3$O$_{7-x}$ (YBCO) single crystals \cite{Tsuchiya2001a}, and in the range 30--70~GHz for temperatures bewteen 60~K and 80~K in nanostructured YBCO with BaZrO$_3$ nanorods \cite{Pompeo2009,Torokhtii2017,Pompeo2018}; low-$T_c$ Nb thin films exhibit depinning frequencies in the 5--20~GHz range \cite{Silva2011,janjusevic}. Although it was reported that very thin films show increased $f_p$, e.g. \cite{janjusevic}, we do not believe that this is the case, since the thickness of our TBCCO films (also as compared to the London penetration depth) is not significantly different from other cuprates investigated \cite{Pompeo2009,Torokhtii2017,Pompeo2018}. We stress that the parameter $r$, hence $f_p$, is not affected by calibrations and geometrical factors: looking at Eq.s \eqref{eq:rhoGR} and \eqref{eq:measrho}, it is easy to see that $r$ is directly measured from $Q$ and $f_{res}$. 

A high depinning frequency is usually considered the necessary condition for the operation of a superconductor at high frequency in  dc magnetic field. However, a more complete analysis shows that for the compound under study this is not sufficient.

In Figure \ref{fig:param}b we report the vortex viscosity $\eta$ as a function of the dc magnetic field. Although the determination of the absolute value of $\eta$ depends on the determination of the film thickness, a 30\% uncertainty on the determination of the thickness of the film (much larger than reasonable) does not significantly change  the results. We note that the vortex viscosity is nearly constant, but it attains extremely low values: at 81.4 K $\eta$ it is of the order of $\sim$ 0.5$\cdot$10$^{-8}$ Nsm$^{-2}$, to be compared e.g. to data in YBCO at 70 K: in single crystals one gets $\sim$ 10$^{-7}$ Nsm$^{-2}$ \cite{Tsuchiya2001a}, and similar values in films \cite{Torokhtii2017}. 

The analysis of the pinning constant $k_p$, Figure \ref{fig:param}c, reveals however that $k_p$ is much smaller than the corresponding values in YBCO at  similar reduced temperatures $T/T_c$  \cite{Tsuchiya2001a,Torokhtii2017,Pompeo2009}. Thus, we are led to the important conclusion that the huge values of $f_p$ do not come from some kind of strong pinning. Instead, recalling that $f_p=k_p/(2\pi\eta)$, we must ascribe this very high value for $f_p$ to an anomalously small value for $\eta$. The reasons for this behavior are not clear at present, and they resides in the very fundamental nature of the superconducting state in TBCCO: the vortex viscosity is a lumped parameter that contains the details of the microscopic processes governing the dissipation in the flux--flow state, and it is unlikely to be controlled by some kind of material engineering. Thus, although a direct measure of the depinning frequency yields huge values, a more complete study shows that the peculiar (albeit not known) microscopic state gives rise to exceedingly high losses even in a moderate dc magnetic field.

\section{Conclusions}
\label{sec:conclusion}
In order to assess the possible use of Tl$_2$Ba$_2$CaCu$_2$O$_{8+x}$ for microwave applications in a dc magnetic field, we have measured the magnetic field dependence of the complex surface impedance in thin TBCCO films. We have found that a rich variety of processes exists, even in our moderate fields. In particular, an extrinsic contribution due to grain boundaries or weak links has been identified at low fields. The intrinsic quasiparticle increase process is much more present than in other HTS, thus increasing significantly the dissipation even at 15~K below $T_c$ in moderate fields (below 1~T). The most important finding resides in the huge depinning frequency. However, by performing a complete analysis with a general model for vortex motion, we have shown that the huge values for the depinning frequency are not due to some kind of exceptional pinning (that would increase the interest for applications). Instead, it results from the combination of a  pinning constant which is weaker than in YBCO, and a vortex viscosity that is more than an order of magnitude smaller than in, e.g., YBCO. The latter finding is very hard to reconcile with the present understanding of HTS. Although it cannot be excluded that the depinning frequency could be increased further by acting on, e.g. artificial defects, it has little effect on the practical applications of TBCCO, since it does not reduce the anomalous dissipation.

%
%

\ifCLASSOPTIONcaptionsoff
  \newpage
\fi

\end{document}